\documentclass[]{elsart}
\usepackage[]{graphicx,amssymb,array}
\begin{document}
\runauthor{Kang, Cudell, Ezhela, Gauron, Kuyanov, Lugovsky, Nicolescu and
Tkachenko}

\begin{frontmatter}

\title{Analytic Amplitude Models \\ for Forward Scattering}

\author[USA]{K. Kang\thanksref{USKO}}, \author[belge]{J. R. Cudell},
\author[russe]{V. V. Ezhela}, \author[LPNHE]{P. Gauron}, \author[russe]{Yu. V.
Kuyanov}, \author[russe]{S. B. Lugovsky}, \author[LPNHE]{B. Nicolescu},
\author[russe]{N. P. Tkachenko}

\address[USA]{Physics Department, Brown University, Providence, RI, U.S.A.}
\address[belge]{Inst. de Physique, B\^at. B5, Univ. de Li\`ege, Sart
Tilman, B4000 Li\`ege, Belgium}
\address[LPNHE]{LPNHE, Universit\'e Pierre et Marie Curie, Tour 12 E3, 4
Place Jussieu, 75252 Paris Cedex 05, France}
\address[russe]{COMPAS group, IHEP, Protvino, Russia}
\thanks[USKO]{Supported in part by the US DoE Grant DE-FG02-91ER4068-Task A, ~ Report No. Brown-HET-1296}

\begin{center}
 (COMPETE Collaboration)

{\small {\it (Presented at the 9th International Conference (Blois Workshop) on Elastic and Diffractive
Scattering, Pruhonice, Czech Republic, 9 - 15 June 2000)}}
\end{center}

\begin{abstract}
We report on fits of a large class of analytic amplitude models for forward
scattering against the comprehensive data for all available reactions.
To differentiate the goodness of the fits of many possible parameterizations to
a large sample of data, we developed and used a set of quantitative indicators
measuring statistical quality of the fits over and beyond the typical criterion
of the \( \chi ^{2} /dof \). These indicators favor models with a universal \(
\log ^2 s \) Pomeron term, which enables one to extend the fit down to \(
\sqrt{s} = 4 \) GeV.
\end{abstract}
\end{frontmatter}

\section{Introduction}
This presentation is based on the results of the recent comprehensive studies
on the fits of  the comprehensive analytic amplitude models for the high energy
forward scattering amplitude against all available data of the total cross
sections and real part of the hadronic amplitudes by  the COMPETE
\cite{Cudell,Nicolescu,QCD6} collaboration. The data set is accumulated and
maintained by some of us (COMPAS) for the PPDS data bases \cite{Cudell00}.

The problem of universal description of forward scattering and rising total
cross section has been with us for over two decades ever since the ISR
experiments. Interest in this topic has been revived by the recent activities
at HERA on deep-inelastic and diffractive scattering and also at LEP on $\gamma
\gamma$ scattering. The rising behavior of total cross sections in energy was
suggested theoretically before the ISR results in connection with the rigorous
proof of  the Pomeranchuk theorem \cite{Pom}.

Theoretically the total cross section is related to the imaginary part of the
forward scattering amplitude. Unitarity and analyticity relates the real part
to the imaginary part of the scattering amplitude. The real part of the forward
amplitude $F^{ab}$  can be obtained from the imaginary part
via the well-known substitution rule $ s \to se^{-i\pi/2} $ or, in an
equivalent way, from the derivative dispersion relations \cite{Kang75}. A slow rise of
total cross sections with energy is possible if the leading Regge trajectory
having  vacuum quantum numbers has an intercept slightly larger than 1, {\it
i.e.}, the intercept of the Pomeron as a simple Reggeon is $\alpha_{P}(0)= 1+
\varepsilon$, $\varepsilon$ being a small positive number, which was proved to
be useful for studying the data at non-asymptotic energies \cite{Land92}.
But as the Pomeron intercept has a bearing on the extrapolation of total cross
section to higher energies, such Pomeron term will violate eventually the
Froissart bound \cite{Frois61} $ \sigma_{tot} \leq c \log ^2 s$, $s$ being the square of the C.M. energy, a consequence of the unitarity and positivity of the
imaginary part of the scattering amplitudes in the Lehmann ellipse. In
addition, the simple Pomeron model does not offer a simple and automatic
extension
to the off-shell particle scattering and in particular to deep
inelastic scattering (DIS). It is generally believed that the BFKL
re-summation of energy logarithms is
relevant to high energy hadron processes but
the BFKL scheme
offers no simple clue for the off-shell extension either.

The Pomeron intercept is also a crucial element in HERA DIS analyses and
provides the starting point at low $x$ and low $Q^{2}$, from which perturbative
QCD evolution can be performed and can be tested. Since the Regge-Pomeron
behavior in $s$ originates from that of the large t-channel scattering angle,
$\cos \theta _t$, in hadron-hadron scattering, the same $ J - $ plane
singularity is affecting the corresponding $(x, Q^2)$ region in DIS at HERA.
Though there is some overlap region of the large $s$ behavior of the soft
Pomeron in hadronic scattering and  the small $x$ behavior from perturbative
QCD evolution,  the study of the singularity structure of forward hadronic
scattering lies mostly outside the realm of perturbative QCD. In pQCD,
one may expect that a higher order effects would unitarize the amplitude and
tame the fierce rise observed at large $Q^{2}$ to something compatible with the
Froissart bound.
 But no one has derived reliably unitarized QCD amplitudes.
Moreover such unitarization will involve necessarily multi-gluon exchanges
between the quarks and therefore will require detailed quark structure,
{\it i.e.,} the hadronic wave functions. In doing so, one is likely to
lose the nice properties that the simple Regge-Pomeron exhibits. It is simply
the reality that no one has the QCD-based understanding and derivation of the
forward hadronic scattering amplitudes.

The basic idea of the analytic amplitude method for the  forward $(t=0)$ hadron
scattering is to treat this non-perturbative domain by implementing the
general principles as much as possible, such as analyticity, unitarity,
crossing-symmetry and positivity of total cross-sections, and by supplementing
the scheme by some well-established strong interaction properties. Good
analytic models must describe not only the desired rising behavior at high
energies but also the correct energy behavior of the cross sections when
extended to  medium energies where a smooth analytic behavior of the amplitude
has set in and the secondary Reggeon contributions are needed.  Such general
approach,
unfortunately, does not provide a unique answer. There can be many models that
satisfy these theoretical criteria. In addition, good models are required to
explain all available forward scattering data of all reactions and their
applicability must be judged on a common ground. It is therefore clear that one
needs to develop and use a common procedure and decision-making criteria to
differentiate the goodness of the fits of the models.  In fact a
decision-making procedure was initiated in Ref. \cite{Cudell97} for $pp$ and
$\bar pp$ scattering, which led us to conclude that the exchange-degenerate
Reggeons were not preferred by the forward scattering data and pursued further
in Refs.~\cite{Cudell00} for the collisions $p^{\mp}~p$,
${\pi}^{\mp}~p$, $K^{\mp}~p$,  $\gamma~p$, and $\gamma~\gamma$, which however
led to the conclusion that the forward data could not discriminate between a
simple Regge-Pomeron fit and asymptotic $\log^ 2 s$ and $\log s$ fits for
$\sqrt{s} \ge 9 $ GeV.
\begin{itemize}
\item We have  therefore developed a set of statistical indicators which
measure quantitatively the goodness of the fits and which complement the
usual $\chi ^2$ criterion. The values of these indicators enable us to
differentiate the diverse analytic amplitude models on the basis of
characteristics of their fits.
\item The data set has been improved somewhat by eliminating some preliminary
data on the $\rho$ parameter and by adding newly published data from SELEX
(${\pi}^- N$ and $\Sigma^- N$ at 600 GeV/c) \cite{Dersch: 2000zg} and
OPAL ($\gamma~\gamma$) \cite{Abbiendi: 2000sz}. Thus new SELEX  data of the
($\Sigma^- p$) collision and the published OPAL data are added to the current
simultaneous scan-fits.  We have however excluded all cosmic data
points
\cite{Baltrusaitis: 1984ka,Honda: 1993kv}  as in the previous studies
\cite{Cudell00} 
because the original numerical Akeno(Agasa) data are not available (only data
read from graph) and there are contradictory
statements concerning the cosmic data points of
both Fly's Eye and Akeno(Agasa) in the literature.
\item Recently a two-component soft Pomeron picture was rediscovered
\cite{Gauron: 2000ri},  in which the first component has intercept 1 and comes
from $C = + 1$ three-gluon exchanges and the second component may be thought of
as a unitarized Pomeron with intercept larger than 1, the well-known BFKL
Pomeron associated with 2-gluon exchanges\cite{Bartels00,Vacca}. The first component can take the
quark counting rule into account while the second one is
responsible for the universal rising of total cross sections with
energy.
\end{itemize}
We see that this procedure changes the picture considerably and can
differentiate the models that can be applied to the energy region as low as
$\sqrt{s} = 4  $ GeV.

\section{Analytic Parameterizations for the Forward Scattering Amplitudes}
The different variants of parameterization can be classified basically into
three exemplary classes depending on the asymptotic behaviors of the cross
section, in the limit \( s\rightarrow \infty  \),
as a constant, as \( \log s \) or as \( \log ^{2}s \). We introduce the
following notations accordingly for the imaginary part of the forward
scattering amplitude: 

\begin{equation}
\label{SIG}
ImF^{ab}= s \sigma _{}^{a_{\mp } b}= {\textrm{R}}^{+ab}(s) {\pm}
{\textrm{R}}^{-ab}(s)+{\textrm{P}}^{ab}+{\textrm{H}}^{ab}(s) ,
\end{equation}
where  the $\pm$ sign in formula corresponds to anti-particle(particle) -
particle collisions,
\noindent \vspace*{3mm}

\( {\textrm{R}}^{\pm ab}(s)=Y^{ab}_{\pm}\cdot
(s/s_{1})^{\alpha _{\pm}},\)

$R_{\pm}$ representing the effective secondary-Reggeon $( (f,a_{2}), 
(\rho,\omega ) )$ contributions to the even(odd)-under-crossing amplitude,
\({\textrm{P}}^{ab}=sZ^{ab} \), and \( {\textrm{H}}^{ab}(s) \)
stands for either
\( {\textrm{E}}^{ab}=X^{ab}(s/s_{1})^{\alpha _{\wp }} \), or 
\({\textrm{L}}^{ab}=s\left( B_{ab}\ln (s/s_{1})+A_{ab}\right)  \), with \(
s_{1}=1 \)  GeV \(^{2} \), or \( {\textrm{L}2}^{ab}=s(B^{ab}\ln
^{2}(s/s_{0})+A_{ab}) \) with an arbitrary scale factor $s_0$.

In our work we combine the constant \( A_{ab} \) with \( Z^{ab} \) 
in the \( P^{ab} \) term and rewrite \({\textrm{P}}^{ab}+{\textrm{L}}^{ab}=s \left(B_{ab} \ln (s/s_{1})+ Z_{ab}\right) \) and  also \({\textrm{P}}^{ab} + {\textrm{L}2}^{ab}=s(Z_{ab} + B_{ab} \ln^2(s/s_{0})) \). Analyticity determines the real part 
of $F^{ab}$ from the
formula (1) via the rule as mentioned already. Most of the analytic amplitude
models proposed in the literature \cite{Kang75,Amaldi: 1980kd,log parametrisations} are some variants of the ${\rm R R} {\rm P} {\rm H}$
parameterization or those constrained further by imposing such conditions as
the degeneracy of the leading $C = {\pm}$ Regge trajectories, the universal
rising $B_{ab} = B$ of total cross sections, the factorization of the Regge
residues of $H^{ab}$,
\( H_{\gamma \gamma }=\delta H_{\gamma p}=\delta ^{2}H_{pp} \),
and the quark counting rule for the residues of the Regge-Pomeron terms in
$H^{ab}$, in the interest of reducing the number of parameters. Models
constrained by such additional conditions are indicated by appropriate symbols
$ d, u, qc$ and $nf$ (in the case of not imposing the factorization of $H^{ab}$)
as supplementary subscript or superscript.

In general there are seven (six) adjustable parameters for each pair of collisions ${\overline a}~b$ and $a~b$ in the analytic parameterizations ${\rm R R} {\rm P} {\rm H}$  where ${\rm H} = {\rm E}, {\rm L2} ~ ({\rm L})$ so that there are 24 or 25 parameters  in total  to adjust the simultaneous
fits for the collisions $p^{\mp}~p$, $\Sigma^{-}~p$,
${\pi}^{\mp}~p$, $K^{\mp}~p$,  $\gamma~p$, and $\gamma~\gamma$.
We considered more than 256
different variants of the analytic amplitudes \cite{web}. We summarize in Table 1 the results  of six representative cases that give a \( \chi ^{2}/dof \) smaller
than 1.5 for all all cross section and $\rho$ data for   \( \sqrt{s}\geq 4 \)
GeV. Because of the large number of points, slight upward deviations of the \( \chi
^{2}/dof \) from 1 would imply a very low confidence level. The area of
applicability of the models, $i.e.,$ the low-energy cut-offs for which
\( \chi ^{2}/dof\leq 1.0 \) are shown with numbers in bold.

{\centering {\small \begin{tabular}{|l|c|c|c|c|c|c|c|}
\cline{2-8}
\multicolumn{1}{l|}{}&
\multicolumn{7}{c|}{ {\small \( \sqrt{s_{min}} \) in GeV and number of data
points }}\\
\hline
{\small \( {\textrm{Model code}\, (\textrm{N}_{par})} \)}&
{\small 4 (742)}&
{\small 5 (648)}&
{\small 6 (569) }&
{\small 7 (498) }&
{\small 8 (453) }&
\multicolumn{1}{c|}{{\small 9 (397) }}&
{\small 10 (329) }\\
\hline
\hline
{\small \( \textrm{RRE}_{nf} \)}&
{\small 1.4 }&
{\small 1.1 }&
{\small 1.1 }&
{\small 1.1 }&
{\small 1.1 }&
{\small 1.0 }&
{\small 1.0}\\
\hline
\hline
{\small \( {\textrm{RRL}_{nf}(19)} \)}&
{\small 1.1 }&
{\small \( \bf 0.97 \)}&
\textbf{\small \( \bf 0.97 \)}&
\textbf{\small 1.0} {\small }&
{\small \( \bf 0.96 \)}&
{\small \( \bf 0.94 \)}&
{\small \( \bf 0.93 \)}\\
\hline
{\small \( {\textrm{RRPL}}(21) \)}&
{\small 1.1 }&
{\small \( \bf 0.98 \)}&
{\small \( \bf 0.98 \)}&
{\small \( \bf 0.99 \)}&
{\small \( \bf 0.94 \)}&
{\small \( \bf 0.93 \)}&
{\small \( \bf 0.91 \)}\\
\hline
\hline
{\small \( {(\textrm{RR})^{d}\textrm{ P$_{nf}$ L}2}(20) \)}&
{\small 1.2 }&
{\small \( \bf 1.0 \)}&
{\small \( \bf 1.0 \)}&
{\small \( \bf 0.99 \)}&
{\small \( \bf 0.94 \)}&
{\small \( \bf 0.93 \)}&
{\small \( \bf 0.92 \)}\\
\hline
{\small \( {\textrm{RRPL}2_{u}}(21) \)}&
{\small 1.1 }&
{\small \( \bf 0.97 \)}&
{\small \( \bf 0.97 \)}&
{\small \( \bf 0.97 \)}&
{\small \( \bf 0.92 \)}&
{\small \( \bf 0.93 \)}&
{\small \( \bf 0.92 \)}\\
\hline
{\small \( {(\textrm{RR})^{d}\textrm{ PL}2_{u}}(17) \)}&
{\small 1.3 }&
{\small 1.0 }&
{\small \( \bf 1.0 \)}&
{\small \( \bf 0.98 \)}&
{\small \( \bf 0.94 \)}&
{\small \( \bf 0.93 \)}&
\textbf{\small \( \bf 0.93 \) }\\
\hline
\end{tabular}}\small \par}

\begin{quote}
\medskip{} {\small Table 1: Six representative models in three classes fitting
all cross
section and \( \rho  \) data down to 5 GeV. Numbers in bold represent
the area of applicability of each model.}
\end{quote}
As can be seen from Table 1, the data are compatible with many possibilities
for $\sqrt{s} \geq 9 $ GeV and cannot differentiate the models at this level,
let alone the nature of the Pomeron. Also it seems that sub-leading
trajectories and other non-asymptotic characteristics do not manifest
themselves. The two classes of logarithmic increases seem to
fit better than simple powers. Also reasonable degeneracy of the leading
Reggeon trajectories can be
implemented only for the class of $PL2$ models having a \( \log ^{2}s \)-type
effective Pomeron.
Such degeneracy is in fact expected to hold in global fits to
the forward scattering data of all hadronic processes that include
$p p$ and $K^{+}p$ scattering, which have exotic s-channel, in view of the
Reggeon-particle duality.

\section{Statistical Indicators Measuring the Quality of the Fits}
To distinguish further the nature of the fits in these
models, we need a statistical procedure. This procedure will enable us to
compare and rank the quality of the analytic amplitude models.

The best known quantity is certainly the \( \chi ^{2}/dof \),
or more precisely the \emph{}confidence level (CL).
However, because Regge theory does not apply in the resonance region,
no model is expected to reproduce the data down to the lowest measured
energy. The energy cutoff $\sqrt{s _{min}}$ in Table 1 is \emph{ad hoc}.
Clearly the \emph{range} of energy over which the model can reproduce the data
with
a \( \chi ^{2}/dof\leq 1.0 \) must be a part of the indicators. Also the
quality of the data varies depending on which quantity
or which process one considers. An unbiased way of taking into account the
quality of the data is to assign a weight to each
process or quantity. Given that this will be done to compare models together,
the weights as determined by the best fit is a reasonable choice for the
weight, $i.e.,$ we introduce \[w_{j}=\min \left( 1,\frac{1}{\chi
_{j}^{2}/nop}\right) \]
where \( j=1, \ldots, \) 9 refers to the process, and we define the
renormalized
\( \chi ^{2}_{R}\equiv \sum _{j}w_{j}\chi ^{2}_{j}\).
The number of parameters in the model should be a consideration too, given the
data sample in the range of applicability.
Finally, if a fit is physical in a given range, then its parameters
must be stable with respect to the sub-part of the range: different
determinations
based on a sub-sample must be compatible. Hence certain criteria for the
\emph{stability} of the fits should be taken into consideration as indicators.

We have developed a set of statistical quantities that enable us
to measure the above features of the fits. All these indicators are
constructed so that the higher their values are the better is the quality
of the data description.

\begin{description}
\item [(1)]\textbf{The} \textbf{Applicability Indicator: } It characterizes
the range of energy which can be fitted by the model with a confidence level
bigger CL \( >50\% \). This range can
in principle be process-dependent, but we consider the simplest case here: 

\begin{equation}
\label{A}
A_{j}^{M}=w_{j}\log (E^{M,high}_{j}/E^{M,low}_{j}),\qquad A^{M}={1\over
N_{sets}}\sum _{j}A^{M}_{j}
\end{equation}
where \( j \) is the multi-index {denoting} the pair (data subset,
observable); \( E^{M,high}_{j} \) is the highest value of the energy
in the area of applicability of the model \( M \) in the data subset
\( j \); \( E^{M,low}_{j} \) is the lowest value of the energy in
the area of applicability of the model \( M \) in the data subset
\( j \), and \( w_{j} \) is the weight determined from the best
fit in the same interval (hence \( w_{j} \) will depend itself on
\( E^{M,high}_{j} \) and \( E^{M,low}_{j} \)). In our case the applicability
indicator takes the form: 

\begin{eqnarray}
A^{M}= {1\over 15} \left( A^{M}_{pp,\sigma }+A^{M}_{{\overline{p}}p,\sigma
}+A^{M}_{\pi ^{+}p,\sigma }+A^{M}_{\pi ^{-}p,\sigma }+A^{M}_{K^{+}p,\sigma
}+A^{M}_{K^{-}p,\sigma }+ A^{M}_{\Sigma ^{-}p,\sigma }\right. \nonumber \\
 &  & \nonumber \\
\left.+A^{M}_{\gamma p,\sigma}+A^{M}_{\gamma \gamma ,\sigma }
 + A^{M}_{pp,\rho}+A^{M}_{{\overline{p}}p,\rho
}+A^{M}_{\pi ^{+}p,\rho }+A^{M}_{\pi ^{-}p,\rho }+A^{M}_{K^{+}p,\rho
}+A^{M}_{K^{-}p,\rho }\right).\nonumber
\end{eqnarray}
\noindent
The fit results in detail show that for
$L$, $PL$ and $PL2$ class of models we obtain rather good fits to all cross sections
starting from $E_{min} = 4$ and to cross section and $\rho$ data from $5 \ {\rm
GeV}$ but in some cases with negative contributions to the total cross
sections from terms corresponding to the exchange of the Pomeron-like
objects in low energy part of the area of applicability as defined
above. This is unphysical: we are forced to add an additional constraint
to the area of applicability and  exclude the low energy part
where at least one collision process has a negative contribution from
the Pomeron-like (asymptotically rising) term.  It turned out  that some models
have an empty area of applicability once this criterion was
imposed.
\item [(2)~Confidence-1~Indicator: ] \[C^{M}_{1}=CL\%\] 
\noindent
where the CL refers to the whole area of applicability of the model $M.$
\item [(3)~Confidence-2~Indicator: ] \[C^{M}_{2}=CL\%\] 
\noindent
where the CL refers to the intersection of the areas of applicability
of all models qualified for the comparison (we choose here \( \sqrt{s}\geq 5 \)
GeV for the fits to the cross sections but without \( \rho  \)-data, and
\( \sqrt{s}\geq 9 \) GeV for the fits to both cross sections and \( \rho  \)
data.
\item [(4)~Rigidity~Indicator 1: ] We propose to measure the rigidity of the
model by the indicator
\begin{equation}
\label{R1}
R^{M}_{1}={N_{dp}^{M}(A)\over {1+N^{M}_{par}}}
\end{equation}
The most rigid model has the highest value of the number of data
points per adjustable parameter.
\item [(5)~Reliability~Indicator 2: ] This indicator characterizes the goodness
of the parameter error matrix.
\begin{equation}
\label{R2}
R^{M}_{2}={2\over N_{par}(N_{par}-1)}\cdot \sum _{i>j=1}^{N}\Theta
(90.0-C^{R}_{ij})
\end{equation}
where \( C^{R}_{ij} \) -- is the correlation matrix element in \( \% \)
calculated in the fit at the low edge of the applicability area. For the
diagonal correlator this indicator is maximal and equals \( 1 \).
\item [(6)~Uniformity~Indicator: ] This indicator measures the variation
of the \( \chi ^{2}/nop \) from bin to bin for some data binning
motivated by physics: 
\begin{equation}\label{U}
U^{M}=\left\{ {1\over N_{sets}}\sum _{j}{1\over 4}\left[ \frac{\chi
_{R}^{2}(t)}{N^{t}_{nop}}-\frac{\chi _{R}^{2}(j)}{N^{j}_{nop}}\right]
^{2}\right\} ^{-1},
\end{equation}
where \( t \) denotes the total area of applicability, \( j \)
is a multi-index  denoting the pair (data set, observable).
In our case we use the calculation of the \( \chi _{R}^{2}/nop \)
for each collision separately, i.e. the sum runs as in the case of
the applicability indicator.
\item {\bf (7)~Stability-1~Indicator: } \\

\begin{equation}
\label{S1}
S^{M}_{1}=\left\{ {1\over N_{steps}N_{par}^{M}}\sum _{steps}\sum
_{ij}(P^{t}-P^{step})_{i}(W^{t}+W^{step})^{-1}_{ij}(P^{t}-P^{step})_{j}\right\}
^{-1}
\end{equation}
where: \( P^{t} \) - vector of parameters values obtained from the
model fit to the whole area of applicability; 
\( P^{step} \) - vector of parameters values obtained from the model
fit to the reduced data set on the \( step \), by which we mean shift in the
low edge of the fit interval to the right by 1
GeV. If there are no steps then \( S^{M}_{1}=0 \) by definition; and
\( W^{t} \) and \( W^{step} \) are the error matrix estimates obtained
from the fits to the total and to the reduced on the step \( s \)
data samples from the domain of applicability.
\item {\bf (8) Stability-2 Indicator: } \\

\begin{equation}
\label{S2}
S^{M}_{2}=\left\{ {1\over 2N_{par}^{M}}\sum _{ij}(P^{t}-P^{t(no\: \rho
)})_{i}(W^{t}+W^{t(no\: \rho )})^{-1}_{ij}(P^{t}-P^{t(no\: \rho )})_{j}\right\}
^{-1}.
\end{equation}
The indicator \( S^{M}_{2} \)
characterizes the reproducibility of the parameters values when fitting
to the reduced data sample and reduced number of observable but with the same
number of adjustable parameters and might be strongly correlated
with the uniformity indicator $U^{M}$. In this case, we fit the whole set of
the model parameters to the
full area of applicability (superscript $t$) and the same set of
parameters but to the data sample without $\rho$-data (superscript
$t(no\ \rho)$).~
\end{description}

To complement the usual $\chi^2$ criterion, we have developed and used these
quality measure indicators to rank the models via
the ``league comparison" between the models in eight disciplines of the quality
indicators with equal weight
$$ I^m_k = (A^m,C^m_1,C^m_2,R^m_1,R^m_2,U^m,S^m_1,S^m_2) $$
where the index $m$ describes the model and index $k$ describes the indicator
type.

With all the calculated components of the indicators, it is easy to assign
score points in each discipline to a given model $M$: 
\begin{equation}\label{PMK}
P^M_k = \sum_{m \neq M}{( 2\Theta(I^M_k-I^m_k) + \delta_{I^M_k,I^m_k})}.
\end{equation}
The rank of models is decided by the total score points: 
\begin{equation}\label{PM}
P^M = \sum_k{P^M_k} = \sum_k \sum_{m \neq M}{( 2\Theta(I^M_k-I^m_k) +
\delta_{I^M_k,I^m_k})}
\end{equation}

The scores of the {\bf ACCRRUSS} league comparison are given in Table 2 for the
five high ranking representative models, out of 21 models that passed the high
$CL$ tests of the fit comparison of the
$\sigma_{tot}(s)$- and $\rho (s)$-data.

{\centering {\small \begin{tabular}{|l||c|c|c|c|c|c|c|c|c|}
\hline
{\small Model Code}&
{\small \( A^{M} \)}&
{\small \( C^{M}_{1} \)}&
{\small \( C^{M}_{2} \)}&
{\small \( U^{M} \)}&
{\small \( R^{M}_{1} \)}&
{\small \( R^{M}_{2} \)}&
{\small \( S^{M}_{1} \)}&
{\small \( S^{M}_{2} \)}&
rank  \( P^{M} \)\\
\hline
\hline
{\small \( {\textrm{RRPL}2_{u}(21)} \)}&
\textbf{2.2}&
{\small 68. }&
\textbf{85.} {\small }&
\textbf{23.} {\small }&
{\small 29. }&
\textbf{0.90}&
{\small 0.22 }&
{\small 0.10 }&
\textbf{222}\\
\hline
{\small \( {(\textrm{RR})^{d}\textrm{ P$_{nf}$L}2(20)} \)}&
\textbf{2.2}&
{\small 50. }&
{\small 82. }&
{\small 18. }&
{\small 31. }&
\textbf{0.90} {\small }&
{\small 0.27 }&
{\small 0.41 }&
178\\
\hline
{\small \( {(\textrm{RR})^{d}\textrm{ PL}2_{u}}(17) \)}&
2.0&
{\small 50. }&
\textbf{83.} {\small }&
{\small 16. }&
\textbf{32.} {\small }&
{\small 0.88 }&
\textbf{0.30} {\small }&
{\small 0.67 }&
174\\
\hline
\hline
{\small \( {\textrm{RRL}_{nf}(19)} \)}&
{\small 1.8}&
\textbf{73.}&
{\small 81. }&
{\small 17. }&
\textbf{32.} {\small }&
{\small 0.78 }&
\textbf{0.29 }&
\textbf{1.3 }&
\textbf{222}\\
\hline
{\small \( {\textrm{RRPL}(21)} \)}&
{\small 1.6}&
\textbf{67.} {\small }&
{\small 82. }&
\textbf{26.} {\small }&
{\small 29. }&
{\small 0.75 }&
{\small 0.21 }&
\textbf{1.1  }&
173\\
\hline
\end{tabular}}\small \par}

{\centering \medskip{} {\small Table 2: Quality indicators in five
representative models fitting all forward data.}\small \par}
{\centering
\par}
\section{Results of the Quality Tests of the Fits}
The clearest outcome of the applicability indicator $A^M$ test is that all
models belonging to the class of a simple Regge-Pomeron are eliminated from the
exclusive group that meets $ \chi ^{2}/dof \le 1 $ for \( \sqrt{s}>5 \) GeV.
The best \( \chi ^{2}/dof \) for these
is 1.12 for RRE\( _{nf} \), which is rejected at the 98\% C.L. as we have a
large number
of data points, 648,  for \( \sqrt{s}>5 \) GeV. However, upon checking
where these values of \( \chi ^{2}/dof \)
come from,  we see that the main difference
comes from fitting the \( \rho  \) parameter data, which is much worse
for the class of RRE than the others, though  no model in all variants can fit
the \( \rho  \) data perfectly and in particular those of \( \pi p \) and \( pp
\).

As we can see from Table 2, the two classes of models having double poles or
triple poles achieve comparable levels of quality, and one cannot
decide which is better with these indicators. Clearly we shall need further
physics arguments to differentiate these two  effective Pomeron.

We found that imposing the Johnson-Treiman-Freund
relation for the cross section differences \( \Delta \sigma (N)=5\Delta \sigma
({\pi }),\Delta \sigma (K)=2\Delta \sigma ({\pi }) \)
never has led to an improvement of the fit and in some case degraded
the fit considerably, though they
produced two parametrisations with fewest parameters.


It turns out that the original cosmic experimental data are best fitted
by our high-rank models quoted in Table 2.

The class  of models with a $\log s$ type Pomeron gives excellent fits to the
soft data without violating the unitarity
and can be extended to deep-inelastic scattering \cite{Desgrolard01}
without any further singularity. But it suffers from several drawbacks: First
of all, the Pomeron
term becomes negative below 9.5 GeV, and the split of the leading
meson trajectories is somewhat bigger than what a normal
duality-breaking estimate or a linear extrapolation of the known resonances
would allow \cite{Desgrolard: 2001sf}. As a result, the Pomeron in
this class of variants is inevitably compromising with the crossing
even Reggeon in the Regge region, effectively
counter-balancing the excessive contribution of the $C = even$ Reggeon and thus taming the medium energy behavior while describing the asymptotic
behavior of the amplitude. Though the quark counting rule seems to be
respected to a very good approximation by this effective Pomeron , $i.e.,$ by
the coefficients of the
$\log s$  and of the constant term, this only reinforces the problem of
negativity as it is very difficult to conceive a non factorizing pole
which would nevertheless respect quark counting.

Finally, we conclude that the best fits are given by the class of $PL2_u$
models that contain a triple pole at \( J=1 \) besides a simple pole with the
intercept exactly 1, which thus produce \( \log ^{2}s \), \( \log s \)
and constant terms in the total cross section. Barring the details of the best
model $RRPL2_u (21)$ and its parameters to
\cite{Cudell,Nicolescu}, the most interesting
properties of this model may be that the constant term respects the quark
counting rule to
a good approximation, whereas the \( \log ^{2}s \) term can be taken
as universal, i.e. independent of the process, as
rediscovered in \cite{Gauron: 2000ri} (see also \cite{log2 very high E}).
The universality of the rising term is expected in the case of the
eikonal unitarisation of a bare Pomeron with the intercept larger
than 1, because the coefficient of the rising term turns out to depend
only on the intercept and slope of the bare Pomeron \cite{FFKT}.
But for the J-plane singularities of double and triple pole types
considered in this paper, the structure of such a singularity
and the origin of its universality is less obvious. Nevertheless,
such a singularity at \( J=1 \) may in fact have a theoretical explanation: 
recently, Bartels, Lipatov and Vacca \cite{Bartels00,Vacca} discovered
that there are, in fact, two types of Pomeron in LLA: besides the
well-known BFKL Pomeron associated with 2-gluon exchanges and with
an intercept bigger than 1, there is a second one associated with
\( C=+1 \) three-gluon exchanges and having an intercept precisely
located at 1. Also the factorization of the rising components of the cross
section, $(H_{\gamma p})^2 \to H_{\gamma \gamma} \times H_{p p}$,  is  well
satisfied by the $PL2$ Pomeron. Furthermore, the degeneracy of
the lower trajectories is respected
to a very good approximation, and the model seems extendible to deep
inelastic scattering \cite{Cudell 2001}. This model also respects
unitarity by construction. Hence this solution is the one that currently meets
all phenomenological and theoretical requirements.

A few remarks as for the future directions are in order: a  remaining problem
in the analysis of the forward data is the difficulty
in adequately fitting the data for the \( \rho  \) parameter in \( pp \)
and in \( \pi ^{+}p \) reactions. While extraction of the \( \rho  \)
data from the measurements of the differential cross sections data
at small \( t \) is a delicate problem, re-analysis of these data
will call for simultaneous fits to the total
cross section data and to the elastic differential cross sections
in the Coulomb-nuclear interference region and in the diffractive
cones and thus an extension of the parametrisations considered here to
the non-forward region. One could also consider a class of analytic
models not incorporated in our fits and ranking procedures, class
in which the rising terms would turn on at some dynamical threshold
\( s_{t} \) (demanding the use of exact dispersion relations), or
add lower trajectories to the existing models. Both approaches would
lead to many extra parameters, and will be the subject of a future
study. Secondly the inclusion of other data may very well allow
one to decide finally amongst the various possibilities. One can go
to deep-inelastic data, but the problem here is that the photon occupies
a special position in Regge theory, and hence the singularities of
DIS amplitudes do not need to be the same as those of hadronic amplitudes.
One can also extend the models to non-forward data and off-diagonal
amplitude such as those of diffractive scattering. Such steps will
involve new parameters associated mainly with form factors, but there
are many data, hence there is the hope that this kind of systematic
study may be generalized. Thirdly it is our intention to develop the ranking
scheme further,
probably along the lines of \cite{Haitun}, and to fine-tune the definition
of indicators, in order that a periodic cross assessments of data
and models be available to the community \cite{web}.

\end{document}